\documentclass[reprint,amsmath,amssymb,aps]{revtex4-2}

\usepackage{graphicx}
\usepackage{dcolumn}
\usepackage{bm}

\usepackage{color}
\usepackage{amsmath}
\usepackage{colortbl}

\usepackage[normalem]{ulem} 

\DeclareMathOperator{\sech}{\textrm{sech}}

\begin{document}

\preprint{APS/123-QED}

\title{Enhanced nonlinear quantum metrology with weakly coupled solitons and particle losses}

\author{Alexander Alodjants}
\affiliation{Institute of Advanced Data Transfer Systems, ITMO University, St. Petersburg, 197101, Russia}

\author{Dmitriy Tsarev}
\affiliation{Institute of Advanced Data Transfer Systems, ITMO University, St. Petersburg, 197101, Russia}

\author{The Vinh Ngo}
\affiliation{Institute of Advanced Data Transfer Systems, ITMO University, St. Petersburg, 197101, Russia}

\author{Ray-Kuang Lee}
\affiliation{Physics Division, National Center for Theoretical Sciences, Taipei 10617, Taiwan}
\affiliation{Institute of Photonics Technologies, National Tsing Hua University, Hsinchu 30013, Taiwan}

\email{alexander\_ap@list.ru}

\date{\today}

\begin{abstract}
The estimation of physical parameters with Heisenberg sensitivity and beyond is one of the crucial problems for current quantum metrology. However, unavoidable lossy effect is commonly believed to be the main obstacle when applying fragile quantum states. To utilize the lossy quantum metrology, we offer an interferometric procedure for phase parameters estimation at the Heisenberg (up to $1/N$) and super-Heisenberg (up to $1/N^3$) scaling levels in the framework of the linear and nonlinear metrology approaches, respectively. The heart of our setup is the novel soliton Josephson Junction~(SJJ) system providing the formation of the quantum probe, i.e, the entangled Fock ($N00N$-like) state, beyond the superfluid-Mott insulator quantum phase transition point. We illustrate that such states are close to the optimal ones even with moderate losses. The enhancement of phase estimation accuracy remains feasible both for the linear and nonlinear metrologies with the SJJs, and allows further improvement for the current experiments performed with atomic condensate solitons with a mesoscopic number of particles. 
\end{abstract}


\maketitle

Modern quantum technologies pose fundamentally new requirements for schemes and procedures of measurement, as well as the subsequent estimation of some physical parameters, in the framework of metrology and sensorics tasks~\cite{Pezze2018, Degen2017}. Traditionally, such schemes are based on the interference of quantum states of light and/or matter waves~\cite{Dowling2015}. In particular, the high accuracy of quantum optical measurements represents an extremely important problem for the detection of gravitational waves, in which quantum states of light interfere in specially designed optical Michelson interferometers~\cite{Aasi, Zhao2020, McCuller2020}. Although the atomic interferometers have been known for about 30 years (see e.g.~\cite{Chu}), current experimental facilities open new perspectives for verification of fundamental physical laws~\cite{Aguilera, Muntinga, Lachmann} as well as for sensorics and metrology in real-world applications~\cite{Bongs, Geiger, Hinton, Bidel}. Typically, such facilities exploit Mach-Zehnder (MZ) interferometers based on ultracold atomic ensembles and Bose-Einstein condensates (BEC)~\cite{Becker, Gross2010, Kok2002, Hauth2014, Tsarev2018, Bongs}. In particular, atomic MZ interferometer presumes certain operations (rotations) on the Bloch sphere performed with effectively two-mode atomic systems, see Fig.~3 in~\cite{Pezze2018}. Experimentally, such a system may be created using atomic Josephson junctions obtained by an optical dipole trap, which provides a double-well potential for condensate atoms~\cite{Gati2005, Albiez2005, Salasnich2011}. Alternatively, it is also possible to use two hyperfine atomic states, linearly coupled by two-photon transition~\cite{Muessel2015}. In both cases, the number of condensate atoms is about few hundreds for preserving the collective behavior of condensate atoms~\cite{Anglin2001}.

The quantum approach in metrology presumes the so-called quantum Cramer-Rao (QCR) bound for the root mean square error $\delta\phi$ of estimation of arbitrary physical parameter $\phi$, i.e., $\delta\phi \geq \frac{1}{\sqrt{\nu} \sqrt{F_Q}}\equiv\frac{\delta\phi_{min}}{\sqrt{\nu}}$, where $\nu$ is the number of experimental runs, $F_Q$ is the quantum Fischer information (QFI)~\cite{Toth2014,Dobrzanski2015}.

Without loss of generality, the measurement and estimation procedure of $\phi$-parameter proposes some state transformation described as $\left|\Psi_{\phi}\right\rangle\equiv\left|\Psi(\phi)\right\rangle =e^{-i\phi (\hat{b}^\dag\hat{b})^k}\left|\Psi\right\rangle$,
where $k$ is a positive integer number, $\left|\Psi\right\rangle$ is an initial state that we prepare for the measurement, and $\hat{b}$ ($\hat{b}^\dag$)
is the annihilation (creation) operator that characterizes the quantum bosonic channel accumulating phase $\varphi\equiv\phi (\hat{b}^\dag\hat{b})^k$.

The best precision of the measurement, 
\begin{equation}\label{K}
\delta\phi_{min}=\frac{1}{N^k},
\end{equation}
can be obtained with an ideal balanced maximally path-entangled two-mode $N00N$-state, $\left|\Psi\right\rangle\equiv\left|N00N\right\rangle = \frac{1}{\sqrt{2}}\left(\left|N\right\rangle_a\left|0\right\rangle_b + \left|0\right\rangle_a\left|N\right\rangle_b\right)$, where $N$ is the total average number of particles. 

The Heisenberg limit (HL) for Eq.~\eqref{K} is obtained in the framework of the linear metrology (LM) approach with $k=1$. The nonlinear metrology (NLM) corresponds to the so-called super-Heisenberg limit (SHL) that may be obtained for Eq.~\eqref{K} at $k\geq2$,~\cite{Campo2017, Napolitano2010, Cheng2014, Luis2007, Luis2009}.

Remarkably, losses play a decisive (destructive) role in the real-world implementation of quantum-metrological schemes based on non-classical states~\cite{Thomas-Peter2011}. The losses especially affect the schemes, where $N00N$-states are proposed to improve the parameters estimation accuracy. As a consequence, an ideal (balanced) $N00N$-state quickly loses the HL advantage in the phase estimation procedure even in the presence of insignificant losses in an interferometer with small particle number $N = 20$, cf.~\cite{Dorner2009}.

Alternatively, entangled Fock states may be more suitable for the quantum metrology purposes in the presence of losses~\cite{Huver2008}. However, obtaining such entangled states with a relatively large number of particles is still an open problem both in theory and experiment. Thus, the strategy to achieve the ultimate precision in the phase estimation within a real-world experiment presumes:

(i) the preparation of the $N00N$-state or some other state that allows for the measurement level close to Eq.~\eqref{K} even in the presence of particle losses, whcih should also be suitable for both linear and nonlinear metrology schemes, and 

(ii) the choices on the appropriate environment ($\phi$) and scheme, which provide the maximal value for power degree $k$. 

The aim of this work is to show how both these two problems addressed above may be solved simultaneously by quantum solitons with a mesoscopic number of particles, even in the presence of losses.

We prove in Sec.~A of Supplementary Materials that, for the estimation of phase parameter $\phi$ in terms of (ii), quantum bright solitons potentially demonstrate a higher accuracy than purely Gaussian states or plane waves in the same Kerr-like medium, see (S.3)  in Supplementary Materials.

\begin{figure}[t]
\center{\includegraphics[width=1\linewidth]{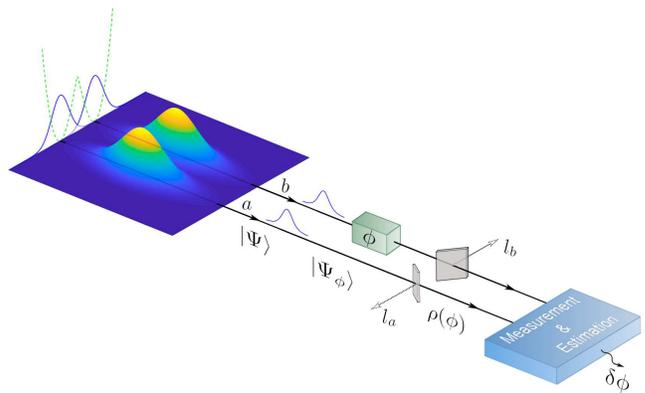}}
\caption{Sketch on three step phase-parameter $\phi$ estimation procedure with quantum soliton Josephson junction~(SJJ) devise as a input state preparation. Possible interaction with environment is provided by two fictitious ``beam splitters" $BS_a$ and $BS_b$ in quantum channels ``a" and ``b", respectively. See text for some details.}
\label{FIG:Scheme}
\end{figure}

The scheme indicating the estimation procedure for phase parameter $\phi$ by quantum bright solitons is plotted in Fig.~\ref{FIG:Scheme}. The procedure includes three important steps. At the first stage we aim at the preparation of entangled Fock ($N00N$-like) state $\left|\Psi\right\rangle$ that provides maximally accessible accuracy for subsequent $\phi$-estimation in the presence of losses. The two entangled modes here correspond to the two arms of MZ-interferometer. At the second stage, a phase-shift $\phi$ between the arms and the losses are accounted. In the framework of NLM it is possible to assume that an additional medium provides a controlled nonlinear phase difference $\phi$ between the solitons, which is accounted at the second stage. The losses for the setup in Fig.~\ref{FIG:Scheme} are introduced at the second stage by two fictitious ``beam splitters" (BSs), which provide a coupling of the setup with the environment~\cite{Dorner2009,Bohmann2015}. At the third final step, the measurement and $\phi$-parameter estimation are performed by including ideal particle detection and outcome estimation~\cite{Thomas-Peter2011}.

Now, let us examine how we can prepare entangled soliton state $\left|\Psi\right\rangle$ to achieve the maximal precision in the phase estimation, $\delta\phi_{min}$. Here, we suggest the quantum soliton Josephson junction~(SJJ) device that represents two weakly coupled bright solitons. We assume that each of the solitons may be described in the framework of the single mode approximation. In this limit, the Hamiltonian for the SJJ-model reads as 
\begin{eqnarray}\label{H_2}
\hat{H}_{SJJ}&=&\kappa N\Bigg\{-\frac{\Lambda}{2}\hat{z}^2 - \frac{1}{2N}\Bigg[\sum_{k=0}^\infty C_{1/2}^k(-1)^k(1-0.21\hat{z}^2)\nonumber \\
&\times&\left(\hat{a}^\dag\hat{b}+\hat{b}^\dag\hat{a}\right)\hat{z}^{2k} + H.C. \Bigg]\Bigg\},
\end{eqnarray}
where $\hat{z}\equiv\frac{1}{N}\left(\hat{b}^\dag\hat{b}-\hat{a}^\dag\hat{a}\right)$; $\hat{a}$ and $\hat{b}$ ($\hat{a}^\dag$ and $\hat{b}^\dag$) are the bosonic annihilation (creation) operators for two solitons modes, see~\cite{Tsarev2020} for details. In Eq.~\eqref{H_2}, $\Lambda\equiv u^2N^2/16\kappa$ is a vital parameter of the SJJ system that includes the nonlinear strength, $u$, coresponding to the two-body atom scattering in the Born approximation, and $\kappa$ characterizing tunneling of particles between the solitons. The analysis of the SJJ system without losses in the general case is performed in Ref.~\cite{Tsarev2020}, where Hamiltonian~\eqref{H_2} clearly demonstrates the quantum ``superfluid-Mott insulator"-like phase transition occurring at some critical value $\Lambda_c$. 

For the lossy quantum metrology scheme, the effect of the coupling with the environment causes some number of particles to remove from each quantum channel introducing extra (vacuum) noises. We model these losses by means of the fictitious BS approach that presumes transmissivities of ``a" and ``b" channels characterized by parameters $\eta_a$ and $\eta_b$ ($\eta_{a,b} \leq1$), respectively. Further, we suppose that channels ``a" and ``b" are physically equivalent and $\eta_a\simeq\eta_b=\eta$. The resulting quantum state in the lossy channel is a mixed state described by the density matrix, cf.~\cite{Dorner2009, Dobrzanski2009}, 
\begin{equation}\label{dens_mat}
\rho = \sum_{l_b=0}^N\sum_{l_a=0}^{N-l_b} p_{l_a,l_b}\left|\xi\right\rangle\left\langle\xi\right|,
\end{equation}
where $\left|\xi\right\rangle = \frac{1}{\sqrt{p_{l_a,l_b}}}\sum_{n=l_b}^{N-l_a}C_{l_a,l_b}^n e^{i n^k\phi}\left|N-n-l_a\right\rangle_a\left|n-l_b\right\rangle_b$ is one of the possible pure states of the system after $l_a$ and $l_b$ particle losses; and $p_{l_a,l_b}= \sum_{n=l_b}^{N-l_a}(C_{l_a,l_b}^n)^2 \equiv \sum_{n=l_b}^{N-l_a}A_n^2 B_{l_a,l_b}^n$ is a normalization factor, see Supplementary Materials. In Eq.~\eqref{dens_mat}, we have also traced out the modes $\left|l_a\right\rangle\left|l_b\right\rangle$.

In the presence of losses, the particle number distributions are broadened, see (S.10-S.16) in Supplementary Materials. Obvious arguments tell us that the broadening of the particle number distribution negatively affects the accuracy of the $\phi$ phase parameter measurement. 
Actually, the uncertainty in the phase parameter estimation occurs due to the quantum-mechanical fluctuations in the particle number difference $\delta\phi\simeq \left\langle(\Delta(N_a-N_b))^2 \right\rangle^{-1/2}$. In this case, we obtain $\delta\phi=[N(1+\eta(N\eta-1))]^{-1/2}$. Without any losses (for $\eta=1$), $\delta\phi$ approaches $1/N$ corresponding to the HL accuracy. 

To be more specific, we examine the corresponding QFI, $F_Q$, in the presence of losses for the setup in Fig.~\ref{FIG:Scheme}. In particular, accounting in Eq.~\eqref{dens_mat} the density matrix representation in diagonal form $\rho = \sum_{i=1}^s\lambda_i\left|\psi_i\right\rangle\left\langle\psi_i\right|$, where $\lambda_i$ and $\left|\psi_i\right\rangle$ are eigenvalues and eigenstates of $\rho$, respectively ($s$ is the rank), we can establish QFI as, cf.~\cite{Jing2014,Braunstein1994} 
\begin{eqnarray}\label{FQ}
F_Q&=&\sum_{i=1}^s\Bigg(\frac{1}{\lambda_i}\left(\frac{\partial\lambda_i}{\partial\phi}\right)^2 \nonumber \\
&+&4\lambda_i\left\langle\psi_i'\Big|\psi_i'\right\rangle - \sum_{j=1}^s \frac{8\lambda_i\lambda_j}{\lambda_i+\lambda_j}\left|\left\langle\psi_i\Big|\psi_j'\right\rangle\right|^2\Bigg),
\end{eqnarray}
where $\left|\psi_i'\right\rangle\equiv\frac{\partial}{\partial\phi}\left|\psi_i\right\rangle$.

Numerical calculation of QFI defined in Eq.~\eqref{FQ} and performed in the case of NLM for large $N$ represents a non-trivial computational task. In this work, we restrict ourselves by studying only the upper bound of the QFI, denoted as $\tilde{F}_Q$ ($F_Q\leq\tilde F_Q$), which reads
\begin{equation}\label{Fisher_Q}
\tilde{F}_Q = 4\left[\sum_{n=0}^Nn^{2k}A_n^2 - \sum_{l_b=0}^N\sum_{l_a=0}^{N-l_b}\frac{\left(\sum_{n=l_b}^{N-l_a}n^kC_{l_a,l_b}^n\,^2\right)^2}{\sum_{n=l_b}^{N-l_a}C_{l_a,l_b}^n\,^2}\right],
\end{equation}
where $A_n$ is $C_{l_a,l_b}^n$ coefficient taken at $l_a,l_b=0$~\cite{Dorner2009, Dobrzanski2009}.

\begin{figure}[t]
\center{\includegraphics[width=0.9\linewidth]{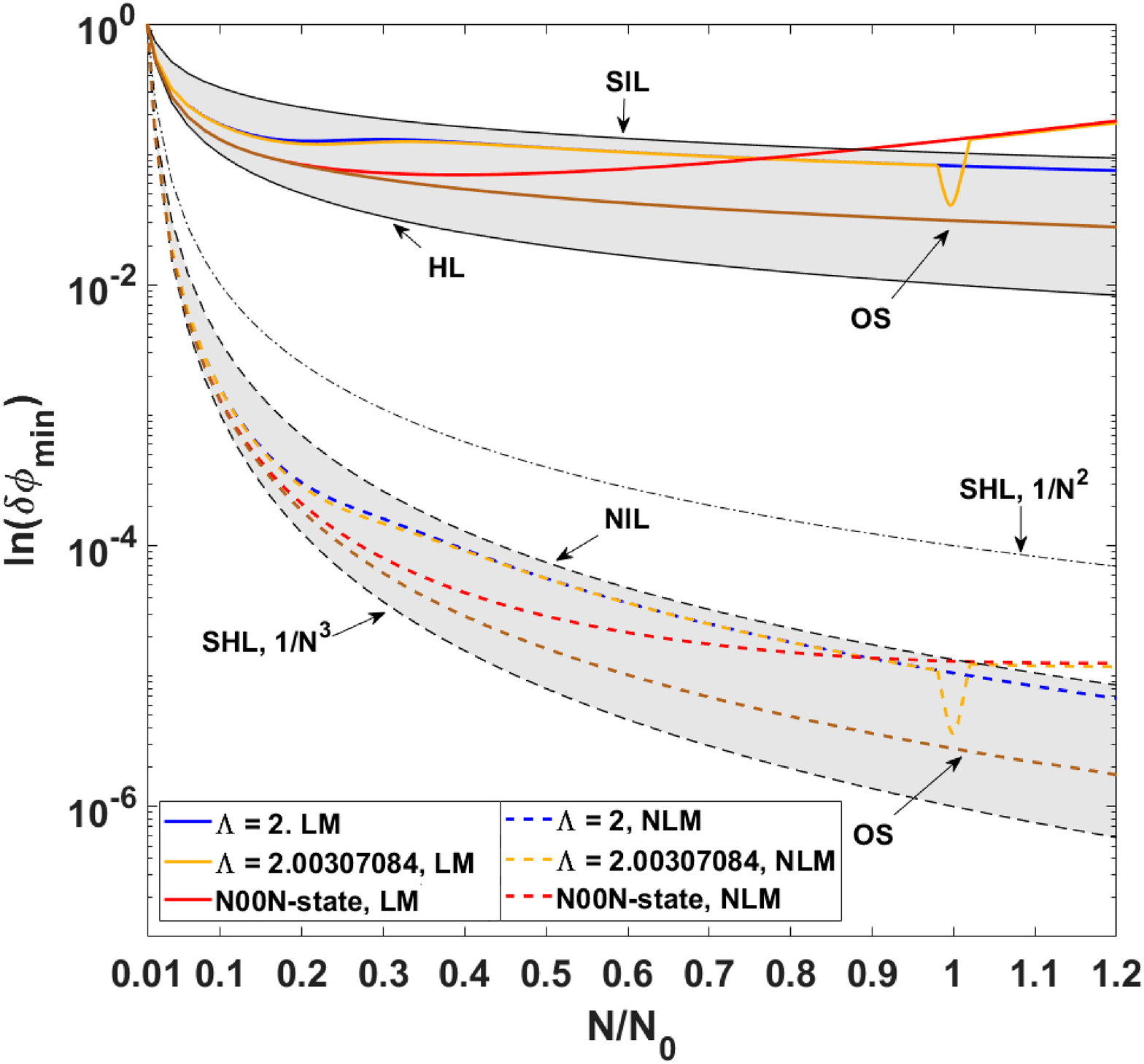}}
\caption{
The logarithmic plot of the best accuracy $\delta\phi_{min}$ of phase estimation based on the SJJ as a function of the normalized number of particles $N/N_0$. The parameters used are $N_0=100$, $\eta_a=\eta_b=0.95$. The upper shadow region depicts the area between the SIL and HL for $k=1$. The lower shadow region corresponds to the area between the NIL and SHL at $k=3$, see Eq.~\eqref{K}. See more detail in text.}
\label{FIG:Fisher_N}
\end{figure}

In Fig.~\ref{FIG:Fisher_N} we establish the minimal error (maximal accuracy) for the phase estimation (represented in the logarithmic scale) as a function of the normalized particle number $N$. The upper (solid) curves correspond to the LM limit obtained at $k=1$ for Eq.~\eqref{H_2}. The nonlinear phase estimation with solitons is established by the lower (dashed) curves plotted at $k=3$. The intermediate (black dashed) curve establishes SHL $\propto 1/N^2$ obtained with $k=2$. These limits are discussed in Ref.~\cite{Luis2009} and correspond to the phase estimation by means of usual Kerr-like medium with ``plane waves". The SJJ environment generates a two-mode Fock state with a particular superposition given by the coefficients $A_n$ that are dependent on $\Lambda$ parameter and defined in the Supplementary Material. For the optimal state (OS, the brown curve in Fig.~\ref{FIG:Fisher_N}), we optimize the numerical QFI $F_Q$~\eqref{Fisher_Q} over all such two-mode Fock state superpositions.

Figure~\ref{FIG:Fisher_N} exhibits two important features, which are relevant to the quantum metrology performed with bright solitons. First, the absolute value of phase estimation accuracy is minimal with the coupled solitons even in the presence of losses and for coherent probes (the blue dashed curve in Fig.~\ref{FIG:Fisher_N}) in comparison to other metrological approaches beyond the HL available now (see the lower shadow region in Fig.~\ref{FIG:Fisher_N} and cf.~\cite{Luis2009}).

Second, the phase estimation with the initially prepared SJJ-system establishes the best accuracy for both linear and nonlinear metrological purposes in comparison to relevant $N00N$-states (see the red curves in Fig.~\ref{FIG:Fisher_N}). For that, in the vicinity of the phase transition point we require $N=N_0$ particles (see the yellow curve in Fig.~\ref{FIG:Fisher_N}). Moreover, the minimal value of the estimated phase parameter is close to the results obtained with the so-called optimal states (OS), which may be found for the lossy metrology by the QFI numerical optimization procedure, cf.~\cite{Dorner2009}. A particular value of $N_0$ is determined from the experimental facilities, which are used for the SJJ-device design. The particle number involved in Fig.~\ref{FIG:Fisher_N} is assumed to be $N_0=100$.

\begin{figure}[ht]
\center{\includegraphics[width=0.9\linewidth]{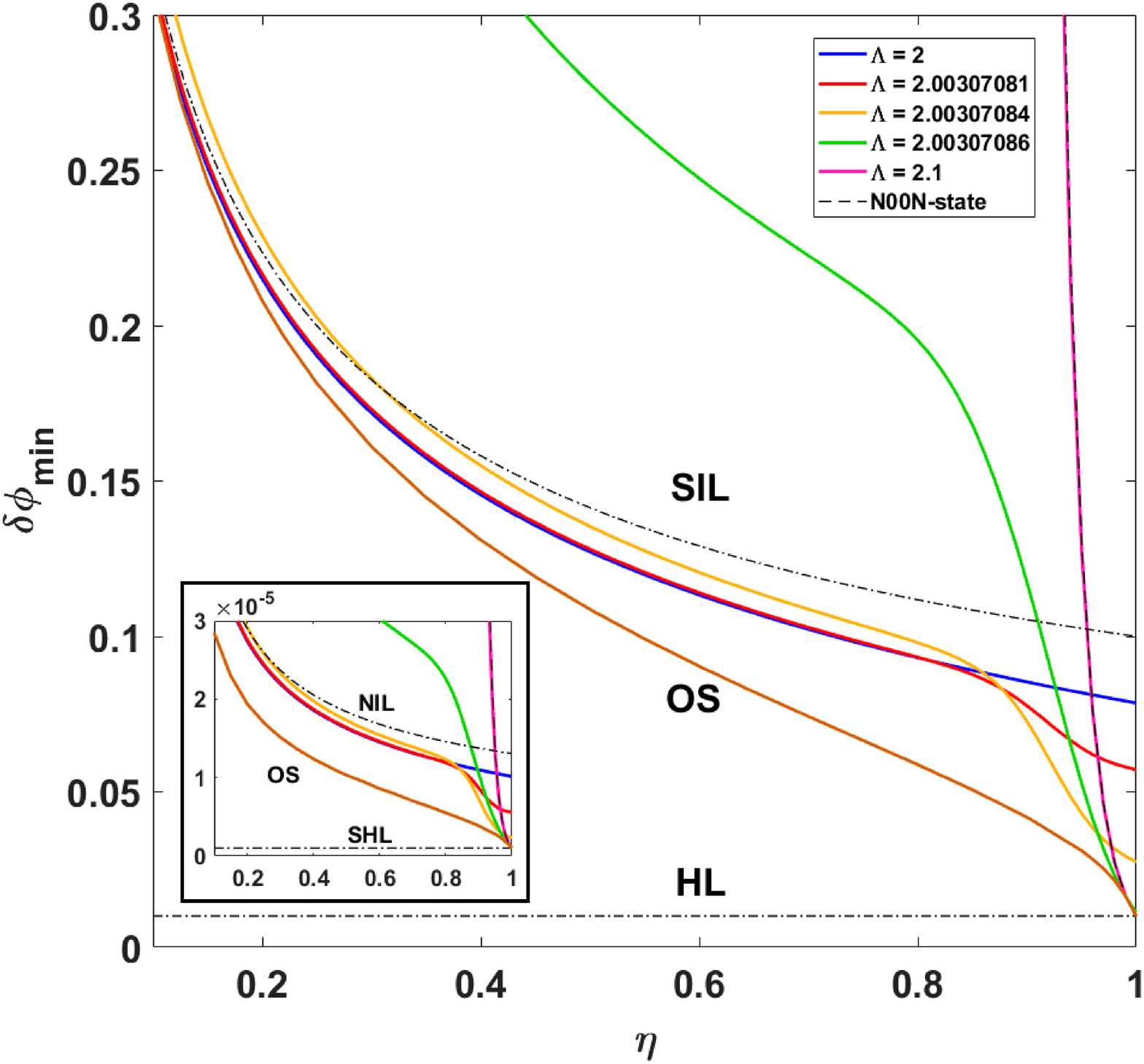}}
\caption{The best accuracy $\delta\phi_{min}$ of the phase estimation based on the SJJ vs. transmissivity $\eta_a=\eta_b\equiv\eta$ of the fictitious BSs for the linear metrology ($k=1$) and nonlinear one~(the inset, $k=3$) with $N=100$ particles. See more detail in text.}
\label{FIG:Fisher}
\end{figure}

The upper black line in Fig.~\ref{FIG:Fisher_N} corresponds to the limiting case, i.e. the standard interferometric limit (SIL), and characterizes the phase estimation procedure with initially prepared coherent atomic wave packets. The SIL reads as $\delta\phi_{SIL}=1/\sqrt{\eta N}$, it can be obtained from Eq.~\eqref{Fisher_Q} at $k=1$ and represents the generalization of the SQL in the presence of particle losses, see Eq.~(S.25) in Supplementary Materials, cf.~\cite{Dobrzanski2009}. Similarly, one can define the nonlinear interferometric limit (NIL), $\delta\phi_{NIL}$, which occurs in the framework of the nonlinear metrology approach. Setting $k=3$ in Eq.~\eqref{Fisher_Q} for the coherent atomic wave packets we can obtain $\delta\phi_{NIL}=1/\sqrt{\eta N^5}$.

In Fig.~\ref{FIG:Fisher}, we represent the accuracy $\delta\phi_{min}={\tilde{F}_Q}^{-1/2}$ versus the channel transmissivity, $\eta$-parameter, for the linear ($k=1$) and nonlinear ($k=3$) metrology phase estimation procedures, respectively. The NLM peculiarities are given in the inset of  Fig.~\ref{FIG:Fisher}.

At $\eta=1$ all the curves start at the points representing the accuracy of the lossless metrology with the SJJs at correspondent $\Lambda$. For the SJJ-device in the Mott-insulator regime, at $\Lambda\geq2.1$, the accuracies of the linear and nonlinear metrology match the HL ($\delta\phi_{min}=1/N$) and SHL ($\delta\phi_{min}=1/N^3$), respectively. In the presence of losses, $\delta\phi_{min}$ grows depending on the value of $\Lambda$-parameter, which is relevant to the performance of the SJJs as a device for probe state preparation in Fig.~\ref{FIG:Scheme}. The curves with various $\Lambda$ in Fig.~\ref{FIG:Fisher} correspond to the crossover region from the superfluid ($\Lambda=2$) to Mott-insulator ($\Lambda=2.1$) regimes that take place in the SJJ system. 

Remarkably, in the LM case the difference between $F_Q$ and $\tilde{F}_Q$ for the SJJ model is more than satisfactory excluding the phase transition point. Only less than 0.5{\%} discrepancy can be seen before the phase transition and less than 4{\%} after the phase transition. At the phase transition point, the difference between $F_Q$ and $\tilde{F}_Q$ reaches 60 {\%}. In the presence of losses it happens in the vicinity of the inflection point with $\eta\lesssim0.9$, see Fig.~\ref{FIG:Fisher}. At this point, $F_Q$ behaves more sharply than $\tilde{F}_Q$. Notice, the plots in Fig.~\ref{FIG:Fisher_N} are represented for $\eta=0.95$, which corresponds to vanishing discrepancy between $F_Q$ and $\tilde{F}_Q$.

From Fig.~\ref{FIG:Fisher} (the red and blue curves), it is clearly seen that at moderate values of $\eta$ the phase estimation accuracy approaches $\delta\phi_{SIL}$ for the SJJ system close to the superfluid regime, $\Lambda < \Lambda_c$. The behaviour of $\delta\phi_{min}$ beyond critical point $\Lambda \geq \Lambda_c$ is determined by the properties of entangled Fock states at the input of the setup in Fig.~\ref{FIG:Scheme} in the presence of losses. Even at $\Lambda=2.1$ (the magenta curves in Fig.~\ref{FIG:Fisher}), accuracy $\delta\phi_{min}$ approaches the one obtained for the ideal $N00N$-state, $\delta\phi_{min}=\delta\phi_{\eta}=1/N^k\eta^{N/2}$, see Eq.~(S.21) in Supplementary Materials. Strictly speaking, use of such states in metrology is justified (for a given particle number $N$) if the transmissivity of quantum channels satisfies condition $\eta \geq \eta_c\equiv e^{-2k/N}$. Notably, in the NLM case with solitons ($k=3$), reasonable value $\eta_c=0.9$ of $\eta$-parameter leads to the mesoscopic number of particles $N\simeq57$ required to achieve a high precision phase estimation. In the LM domain~\cite{Dobrzanski2009} the same particle number requires essentially higher $\eta_c\simeq0.97$ transmissivity of the quantum channels. 

However, in the presence of significant losses ($\eta<<\eta_c$) the entangled Fock or $N00N$-states are not applicable, and the spin-squeezed states with $\Lambda<\Lambda_c$ demonstrate the accuracy better than the SIL. The best accuracy for $\eta<0.9$ is provided by the borderline state with $\Lambda=\Lambda_c$ due to the existence of planar spin-squeezing in the entangled Fock modes (cf. Ref.~\cite{Tsarev2020}), see the yellow curve in Fig.~\ref{FIG:Fisher}.

In Fig.~\ref{FIG:Fisher}, we also compare the results with the OS (the brown curves). As seen, the results with the OS and the states provided by the SJJ are close to each other in the regions where $0<\eta\leq0.3$ and $0.9<\eta\leq1$, which correspond to the high and low losses limits, respectively. The difference between the optimal and SJJ states is notable at intermediate values of $\eta$ close to inflection points, see Fig.~\ref{FIG:Fisher}. Such a behaviour takes place due to the phase-transition effect that occurs abruptly and evokes a rapid quantum state transformation.

Let us estimate $\eta_c$ parameter for the current experiments with atomic solitons observed in $\mathrm{{}^7Li}$ condensate possessing $N=5000$ number of atoms, cf.~\cite{Kevrekidis2008, Strecker2002, Khaykovich2002, Nguyen2014}. To be more specific, we discuss here one-body losses, which correspond to exponential decay law $e^{-\gamma t}$ of the condensate particle number within the rate window $\gamma\simeq 0.1 \div 0.01$ $s^{-1}$. Obviously, the critical value of parameter $\eta_c$ corresponds to critical time scale $t_c=2k/\gamma N$ that limits the observation of the phase estimation with a soliton $N00N$-like state. The maximal value of $t_c$ may be obtained for the NLM setup, $t_c=12 \div 120$ $ms$ ($k=3$); it is comparable with the typical atomic soliton observation and soliton collision time scales in the experiments~\cite{Kevrekidis2008, Strecker2002, Khaykovich2002, Nguyen2014}. However, we expect sufficient improvement in the measurement and phase estimation procedure at the time scales obeying condition $t<<t_c$. The enhancement of $t_c$ represents a non-trivial experimental task; it requires essential reduction of particle number $N$, cf.~\cite{Tsarev2020}. At the same time, we must increase parameter $u$ to keep the value of $\Lambda$ at a certain level providing the bright soliton formation and interaction beyond the phase transition point, see Fig.~\ref{FIG:Fisher}. The latter may be achieved with the improvement of the Feshbach resonance technique~\cite{Chin2010}. We hope that the observation of the SJJ system with mesoscopic number of atoms ($N\simeq100\div1000$) will represent the core for new generation experiments with atomic solitons. 

In conclusion, by utilizing the soliton Josephson Junction system as a quantum probe, we propose the interferometric procedure for the appropriate phase parameters estimation at the Heisenberg (up to $1/N$) and practically unique super-Heisenberg (up to $1/N^3$) scaling levels in the framework of linear and nonlinear metrology approaches, respectively. Counter-intuitively, operating near the quantum phase transition point helps to sustain the accuracy of the phase estimation even in the presence of particle losses. With the QFI, we reveal main features of such a quantum metrology. Our results contribute to further improvement of the current experiments performed with atomic condensate solitons containing a mesoscopic number of particles. 

\section*{Acknowledgement}
This work was financially supported by the Grant of RFBR, No 19-52-52012. RKL was supported by the Ministry of Science and Technology of Taiwan (No 108-2923-M-007-001-MY3. and No 109-2112-M-007-019-MY3).


\preprint{APS/123-QED}

\section*{Supplementary materials: Enhanced nonlinear quantum metrology with weakly coupled solitons and particle losses}

\subsection{Quantum bright solitons as a tool for nonlinear metrology applications in the single mode approximation }

Let us prove that the quantum bright solitons provide the best accuracy in terms of (ii). We assume that the medium ($\phi$) supports the formation of the bright matter-wave soliton described by wave function
\begin{equation*}
\psi(x,t) = \frac{N\sqrt{u}}{2}\sech\left[\frac{Nux}{2}\right]e^{i\frac{N^2u^2}{8}t},\eqno(S.1) 
\end{equation*}
where $u$ characterizes the Kerr-like nonlinearity of the medium. Now, we have $\psi(x,t)$ obeying the normalization condition, $\int|\psi|^2dx = N$, where $N=\langle\hat{N}\rangle$ is the average particle number. If the particle number is not too large, we are able to examine the single (quantum) mode approximation typically used for Gaussian wave packets~[S1, S2, 24].

Consider a classical Hamilton function 
\begin{equation*}
H = \int dx \psi^{*}(x,t)\left(-\frac{1}{2}\frac{\partial^2}{\partial x^2} - \frac{u}{2}|\psi(x,t)|^{2}\right)\psi(x,t),\eqno(S.2) 
\end{equation*}
where $\psi(x,t)$ is the wave function ansatz of the bright soliton with the dimensionless variables $x$ and $t$. Substituting $\psi(x,t)$ into $H$, one can obtain $H = -\frac{u^2}{24}N^3$.

The quantum version of the Hamilton function is the Hamiltonian operator that reads as 
\begin{equation*}
\hat{H}_{\phi} = \phi\hat{N}^3\equiv
\phi(\hat{b}^{\dag}\hat{b})^3,\eqno(S.3) 
\end{equation*}
where $\phi~\equiv-\frac{u^2}{24}$ is the phase parameter suitable for the estimation procedure. Operators $\hat{b}$ and $\hat{b}^{\dag}$ are the annihilation and creation operators characterizing soliton quantum properties in the single mode approximation, see Fig.~1.

Thus, from Eq.~(S.3) it is clear that instead of $k=2$, which is valid for Gaussian states, the implementation of the quantum solitons in the setup of Fig.~1 provides the maximally accessible Kerr-like phase shift in the medium possessing degree $k=3$ in $\hat{H}$ in respect to the particle number $\hat{N}=\hat{b}^{\dag}\hat{b}$.

\subsection{Quantum SJJ-model as a tool for state preparation}

In this work we consider the soliton Josephson Junction (SJJ) model introduced in Ref.~[36]. A fully quantum description of the SJJ is given by the Hamiltonian operator in Eq.~(2). We establish initial state $|\Psi\rangle$ of the SJJ in two-mode Fock basis $|N-n\rangle_a|n\rangle_b$ as
\begin{equation*}
|\Psi\rangle = \sum_{n=0}^N A_n|N-n\rangle_a|n\rangle_b,\eqno(S.4) 
\end{equation*}
where time-dependent coefficients $A_n$ fulfill stationary Schr{\"o}dinger equation 
\begin{equation*}
i\frac{d A_n(\tau)}{d\tau} = \langle N-n,n|\hat{H}|\Psi(\tau)\rangle \eqno(S.5)
\end{equation*}
and obey normalization condition $\sum_{n=0}^N |A_n|^2=1$; throughout this work we use effective dimensionless time $\tau=\kappa N t$. 

Substituting Eq.~(2) into Eq.~(S.4) for unknown coefficients $A_n(\tau)$ we obtain 
\begin{equation*}
i\dot{A}_n = \alpha_n A_n + \beta_nA_{n+1} +\beta_{n-1}A_{n-1},\eqno(S.6)
\end{equation*}
where we introduce the following notations:
\begin{equation*}
\alpha_n = -\frac{\Lambda}{2}\left(\frac{2n}{N}-1\right)^2, \nonumber
\end{equation*}
\begin{widetext}
\begin{equation*}
\beta_n = -\frac{1}{N^2}\Bigg(\left[1-0.21\left(\frac{2n}{N}-1\right)^2\right](n+1)\sqrt{(N-n)(N-n-1)}
+\left[1-0.21\left(\frac{2(n+1)}{N}-1\right)^2\right](N-n)\sqrt{n(n+1)}\Bigg).\nonumber\eqno(S.7)
\end{equation*}
\end{widetext}

In this work, by following Ref.~[36], we find $A_n$ numerically, by considering stationary solution of Eq.~(S.6) with Eq.~(S.7), which is $A_n(\tau) = A_n e^{- iE_n\tau}$, where $E_n$ specifies the eigenenergy spectrum for Hamiltonian~(2).

Practical implementation of the proposed SJJ-model is evident from Fig.~1. Two tunnel-coupled bright solitons with mesoscopic number of particles ($N=5000$ and less) can be achieved within the atomic optics domain~[40-43]. In this case we can speak about matter-wave solitons of atomic BEC possessing negative scattering length. 

On the other hand, tunnel-coupled bright solitons may be obtained in photonics by means of optical pulse propagation in narrow band-gap semiconductor materials, see e.g.~[S3]. The recent progress achieved with exciton polariton solitons in planar AlGaAs waveguides with quantum wells allows to speak about the creation of an on-chip SJJ device based on the photonics platform, cf.~[S4].

In the current paper, we study the atomic BEC soliton platform. In Eqs.~(2),~(S.4)-(S.7) we use dimension-less parameter $u=2\pi|a_{sc}|/a_{\perp}$ that characterizes Kerr-like (focusing) nonlinearity, $a_{sc}<0$ is the s-wave scattering length for attractive particles, $a_{\perp}$ is the characteristic trap scale, and $m$ is the particle mass, cf.~[36]. Notably, critical particle number $N_c$, at which the condensate collapses, for $\mathrm{{}^7Li}$ BEC solitons is $N_c = 0.67 a_{\perp}/|a_{sc}|$ implying $5.2\times 10^3$ particles in the soliton,~[42]. Thereby, $N_c$ represents the upper physical bound for the particle number that limits the quantum metrology scheme in Fig.~1 with atomic solitons possessing negative scattering length. 

\medskip 

\subsection{Soliton-based Interferometer with losses}

Consider the scheme of quantum metrology plotted in Fig.~1 in the presence of particle losses. For our purposes we take state~(S.4) as initial and then take into account the losses in two arms of the interferometer via the fictitious BS approach. This approach represents a powerful tool for modelling the coupling of quantum macroscopic superposition states with the environment. After two BSs the ``input" two-mode Fock state in~(S.4) transforms into (cf.~[39]) 
\begin{widetext}
\begin{equation*} 
\left|N-n\right\rangle_a\left|n\right\rangle_b\rightarrow \sum_{l_b=0}^N\sum_{l_a=0}^{N-l_b}\sqrt{B_{l_a,l_b}^n}\left|N-n-l_a\right\rangle_a\left|n-l_b\right\rangle_b\left|l_a\right\rangle\left|l_b\right\rangle,\eqno(S.8)
\end{equation*}
\end{widetext}
where $l_{a}$ and $l_{b}$ are the numbers of particles lost from ``a" and ``b" channels, respectively. We have no interest in the lost particles here, so further we trace out states $\left|l_a\right\rangle\left|l_b\right\rangle$ for simplicity. In~(S.8) we also introduce coefficient 
\begin{equation*}
B_{l_a,l_b}^n = \left(
\begin{array}{c}
N-n\\
l_a
\end{array}
\right)\left(
\begin{array}{c}
n\\
l_b
\end{array}
\right)
\eta_a^{N-n}(\eta_a^{-1}-1)^{l_a}\eta_b^n(\eta_b^{-1}-1)^{l_b},\nonumber\eqno(S.9) 
\end{equation*}
where $\eta_a$ and $\eta_b$ ($\eta_{a,b} \leq1$) are the transmissivities of BSs in channels ``a" and ``b", respectively. We examine the physically identical arms ``a" and ``b" of the interferometer setting $\eta\equiv\eta_{a}\simeq\eta_{b}$. 

To take into account the phase-shift $\phi$ in Fig.~1 we apply transformation $\hat{U}_\phi$ to state~(S.4), which leads to replacement $\left|N-n\right\rangle_a\left|n\right\rangle_b\rightarrow e^{i n^k \phi}\left|N-n\right\rangle_a\left|n\right\rangle_b$. Then, we can represent density matrix $\rho$ for the final state after particle losses as it is given in Eq.~(3), cf. Fig.~1.

At first, let us examine the influence of particle losses occurring in the interferometer by considering the ideal $N00N$-state 
\begin{equation*} 
\left|N00N\right\rangle = \left(\left|N,0\right\rangle + \left|0,N\right\rangle\right)/\sqrt{2}, \nonumber\eqno(S.10)
\end{equation*}
at the input of the interferometer in Fig.~1. It occurs in the SJJ system without losses far beyond critical point $\Lambda_c$~[36]. In this limit, we suppose formally $k=0$ in Eq.~(3); the coefficients approach to

\begin{equation*}
A_n = 
\begin{cases}
\frac{1}{\sqrt{2}}\textrm{, if }n=0,N;\\
0\textrm{, if }0<n<N. 
\end{cases}
\nonumber\eqno(S.11)
\end{equation*}

Substituting Eq.~(S.11) into Eq.~(3), one can obtain the density matrix for the $N00N$-state after losses as
\begin{widetext}
\begin{equation*} 
 \rho = \eta^N\left|N00N\right\rangle\left\langle N00N\right| + \frac{1}{2}\sum_{l_a=1}^N B_{l_a,0}^0\left|N-l_a,0\right\rangle\left\langle N-l_a,0\right| + \frac{1}{2}\sum_{l_b=1}^N B_{0,l_b}^N\left|0,N-l_b\right\rangle\left\langle 0,N-l_b\right|. 
 \nonumber\eqno(S.12)
 \end{equation*}
Then, taking into account Eq.~(S.9) one can write
\begin{equation*} 
 \rho = \eta^N\left|N00N\right\rangle\left\langle N00N\right| + \frac{1}{2}\sum_{n_a=0}^{N-1} p_{n_a}\left|n_a,0\right\rangle\left\langle n_a,0\right| + \frac{1}{2}\sum_{n_b=0}^{N-1} p_{n_b}\left|0,n_b\right\rangle\left\langle 0,n_b\right|,\nonumber\eqno(S.13)
 \end{equation*}
\end{widetext}
where $n_a = N-l_a$ and $n_b = N-l_b$ are the numbers of particles remained in the arms of the interferometer and 
\begin{equation*} 
p_n = \left(
\begin{array}{c}
N\\
n
\end{array}
\right)\eta^n(1 - \eta)^{N-n}\nonumber\eqno(S.14)
\end{equation*}
is a binomial distribution function of the density matrix diagonal elements.

As seen from~(S.13), the particle losses transform the $N00N$-state into a two-mode Fock states mixture with binomial distribution~(S.14). Notice, in~(S.13) only the term with $n=N$ (the case when no particles are lost) is maximally path-entangled. For $\eta\rightarrow1$ and $N>>1$ and finite $N(1-\eta)$, binomial distribution~(S.14) can be approximated by the Poissonian one 
\begin{equation*}
p_n^{Poisson} = \frac{\left(N(1-\eta)\right)^{N-n}}{(N-n)!}e^{-N(1-\eta)}.\nonumber\eqno(S.15)
\end{equation*}
Finally, at $N(1-\eta)>>1$ one can use the approximation of~(S.15) by the Gaussian distribution 
\begin{equation*}
p_n^{Gauss} = \frac{1}{\sqrt{2\pi N(1-\eta)}}e^{-\frac{(n-N\eta)^2}{2N(1-\eta)}}.\nonumber\eqno(S.16)
\end{equation*}
The distribution in Eq.~(S.16) possesses width $\sigma = 2\sqrt{N(1-\eta)}$ with mean particle number $\bar{n}=N\eta$. For example, if $\eta=0.8$, then $\bar{n}=80$ and $\sigma\approx9$.

\subsection{Phase-estimation bounds with quantum solitons}

In the presence of phase-shift $\phi$ in the scheme shown in Fig.~1, the QFI for pure states $\rho = \left|\Psi(\phi)\right\rangle\left\langle\Psi(\phi)\right|$ is defined as 
\begin{equation*} 
F_Q = 4\left[\left\langle\Psi'(\phi)\Big|\Psi'(\phi)\right\rangle - \left|\left\langle\Psi'(\phi)\Big|\Psi(\phi)\right\rangle\right|^2\right], \nonumber\eqno(S.17)
\end{equation*}
where $\left|\Psi'(\phi)\right\rangle = \frac{\partial}{\partial\phi}\left|\Psi(\phi)\right\rangle$. For mixed state~(3) it can be estimated as 
\begin{equation*} 
F_Q\leq\tilde{F}_Q = \sum_{l_b=0}^N\sum_{l_a=0}^{N-l_b} p_{l_a,l_b}F_Q\left[\left|\xi(\phi)\right\rangle\left\langle\xi(\phi)\right|\right],\nonumber\eqno(S.18)
\end{equation*}
where $F_Q\left[\left|\xi(\phi)\right\rangle\left\langle\xi(\phi)\right|\right]$ is~(S.17) with the state defined as (cf.~(3)) 
\begin{equation*} 
\left|\xi\right\rangle = \frac{1}{\sqrt{p_{l_a,l_b}}}\sum_{n=l_b}^{N-l_a}C_{l_a,l_b}^n e^{i n^k\phi}\left|N-n-l_a\right\rangle_a\left|n-l_b\right\rangle_b. \nonumber\eqno(S.19)
\end{equation*}

Notice, for pure states used in (S.18) we have $F_Q=\tilde{F}_Q$. 

Then, substituting Eq.~(S.19) into Eq.~(S.18) for the upper bound of the QFI $\tilde{F}_Q$ we obtain Eq.~(5). The corresponding density matrix given in Eq.~(3) for the input $N00N$-state has the form
\begin{widetext}
\begin{equation*}
\rho(\phi) = \eta^N\left|N00N\right\rangle\left\langle N00N\right| + \frac{1}{2}\sum_{n=0}^{N-1}p_n\left[\left|n,0\right\rangle\left\langle n,0\right| + \left|0,n\right\rangle\left\langle 0,n\right|\right],\eqno(S.20)
\end{equation*}
\end{widetext}
where $\left|N00N\right\rangle = \left(\left|N,0\right\rangle + e^{i n^k\phi}\left|0,N\right\rangle\right)/\sqrt{2}$ and $p_n$ obeys~(S.14). 

The first term in~(S.20) contains the off-diagonal elements carrying the information about the $\phi$-parameter. At the same time, the sum in~(S.20) consists only of the main-diagonal elements. This occurs due to the initially maximal entangled $N00N$-state collapsing into a Fock-state, when a single particle is lost, c.f.~[35]. QFI upper bound~(5) in this case reads as
\begin{equation*} 
\tilde{F}_Q = N^{2k}\eta^{N}.\nonumber\eqno(S.21)
\end{equation*}
Equation~(S.21) allows to estimate the initial total particle number, $N_{min}$, which provides the minimal error for the $\phi$-measurement with the $N00N$-state in the presence of losses. Such precision requires the maximal value of $\tilde{F}_Q$ that we can find form condition $\partial \tilde{F}_Q/\partial N = 0$. In this case from~(S.21) we obtain the equation for $N_{min}$:
\begin{equation*} 
N_{min} = -\frac{2k}{\ln{\eta}}.\nonumber\eqno(S.22)
\end{equation*}
As example, for the lossy interferometer with $\eta=0.9$, Eq.~(S.22) provides limitations for the particle number $N_{min}\simeq 19$ and $N_{min}\simeq 57$ for $k=1$ and $k=3$, respectively, cf.~[33]. 

Eq.~(S.21) also provides the precision of the phase estimation:
\begin{equation*}\label{ult}
\delta\phi_{\eta}=\frac{1}{\sqrt{\eta^N}N^k}\eqno(S.23)
\end{equation*}
with the input $N00N$-state in the presence of losses. As seen from~(S.23), at $\eta\rightarrow1$, $\delta\phi_{\eta}$ reaches the HL, $\delta_{HL}=1/N$, and SHL, $\delta_{SHL}=1/N^3$, for $k=1$ and $k=3$, respectively. 

On the other hand, the SIL and NIL for the setup in Fig.~1 may be obtained numerically by means of the binomial distributed initial state
\begin{equation*}
\left|\Psi\right\rangle = \frac{1}{\sqrt{2^N}}\sum_{n=0}^N\sqrt{\left(
\begin{array}{c}
N\\
n
\end{array}
\right)}e^{i\phi n^k}|N-n\rangle_a|n\rangle_b
\eqno(S.24)
\end{equation*}
In the presence of losses for~(S.24) one can obtain 
\begin{equation*}
\delta\phi_{k}\propto1/\sqrt{\eta} N^{k-1/2}. \eqno(S.25)
\end{equation*}
The numerical simulations reveal that for $k=1$ Eq.~(S.25) matches the SIL, $\delta\phi_{1}=\delta\phi_{SIL}\equiv1/\sqrt{\eta N}$, cf.~[31]. On the other hand, for $k\geq1$~(S.24) provide the scaling of NIL, for example $\delta\phi_{3}\propto\delta\phi_{NIL}\equiv1/\sqrt{\eta N^5}$.

\renewcommand{\bibnumfmt}[1]{S#1.\hfill}

\end{document}